# Phase coherent transport in bilayer and trilayer MoS$_2$

Leiqiang Chu,[1,2*] Indra Yudhistira,[1,3*] Hennrik Schmidt,[1,3] Tsz Chun Wu,[5] Shaffique Adam,[1,3,6,†] and Goki Eda[1,3,4,‡]

[1]*Department of Physics, National University of Singapore, 3 Science Drive 3, Singapore 117543*

[2]*Department of Physics, Shaoxing University, Shaoxing, China, 312000*

[3]*Centre for Advanced 2D Materials and Graphene Research Centre, National University of Singapore, 6 Science Drive 2, Singapore 117546*

[4]*Department of Chemistry, National University of Singapore, 2 Science Drive 3, Singapore 117551*

[5]*Department of Physics and Astronomy, Rice University, Houston, Texas 77005, USA*

[6]*Yale-NUS College, 6 College Ave West, Singapore 138614*

**Abstract:**

Bilayer MoS$_2$ is a centrosymmetric semiconductor with degenerate spin states in the six valleys at the corners of the Brillouin zone. It has been proposed that breaking of this inversion symmetry by an out-of-plane electric field breaks this degeneracy, allowing for spin and valley lifetimes to be manipulated electrically in bilayer MoS$_2$ with an electric field. In this work, we report phase-coherent transport properties of double-gated mono-, bi-, and tri-layer MoS$_2$. We observe a similar crossover from weak localization to weak anti-localization, from which we extract the spin relaxation time as a function of both electric field and temperature. We find that the spin relaxation time is inversely proportional to momentum relaxation time, indicating that D'yakonov-Perel mechanism is dominant in all devices despite its centrosymmetry. Further, we found no evidence of electric-field induced changes in spin-orbit coupling strength. This suggests that the interlayer coupling is sufficiently weak and that electron-doped dichalcogenide multilayers behave electrically as decoupled monolayers.

[*] L. C. and I. Y. contributed equally to this work.
Email: g.eda@nus.edu.sg

## I. INTRODUCTION

Group 6 layered transition metal dichacogenides (TMDs) emerged as promising candidates for spintronics and valleytronics due to their unique crystal structures and strong spin-orbit interaction [1,2]. Since they are centrosymmetric for even number of layers and non-centrosymmetric for odd number of layers, they provide a unique platform for systematically studying spin and valley physics. Understanding the interplay between these quantum degrees of freedom is key to developing novel device concepts [3,4]. The spin-valley coupling effect has been widely investigated by optically circularly polarized excitation in monolayers, which makes the valley polarization possible [5,6]. In bilayer TMDs, the inversion symmetry can be broken by an external out-of-plane electric field, allowing electrical control of their physical properties [7,8]. Applying an out-of-plane electric field in bilayer TMDs breaks valley degeneracy and recovers spin-valley locking that is unique in odd-layer materials. Correspondingly, the spin and valley lifetime dynamics is predicted to change with application of electric and magnetic field [9]. Using electric field to restore the spin-valley coupling in optical-electrical measurement has been demonstrated in device concepts [10,11]. Meanwhile, electric-field tunability of quantum interference effect in multi-layer and bulk TMDs has been realized experimentally through quantum interference effect in magneto-transport study with ionic liquid gating in the metallic state [12-14].

The effective electrical control of spin-orbit coupling strength is of great interest to the realization of spin- and valleytronic devices. However, field-induced effects in multi-layer TMDs are multifold and complex, and their impact on the experimentally measurable quantities remains elusive. Though quantum transport studies in multilayer by ionic gating manifest valley-contrasting properties, charge carriers also occupy the other valleys such as $\Lambda$ point between the $\Gamma$ and $K$ points, which could inhibit the spin relaxation dynamics. However, for few-layer system such as in 1-3L MoS$_2$, the band minimum could lie in the $K$-point [15]. The reduced effect from other valleys could uncover the spin relaxation dynamics with symmetry argument as a function of thickness. Here, we present a comprehensive theoretical and experimental studies on the effect of out-of-plane electric field on spin-orbit interaction strength in electron-doped mono-, bi-, and trilayer MoS$_2$ through phase-coherent quantum transport. Our findings show that a significantly large electric field is needed to achieve sizeable changes in spin-orbit coupling strength due to weak interlayer coupling of the conduction band electrons.

## II. RESULTS AND DISCUSSION

We start with our theoretical modelling of the electric field-induced effect. The tunability of Zeeman-type spin splitting in bilayer TMDs with broken inversion symmetry are sensitive to two factors: conduction (valence) band interlayer hopping strength for electrons (holes) [9] and Bychkov-Rashba (BR) spin-orbit coupling strength [16]. BR spin-orbit coupling causes the vertically applied electric field in inversion asymmetric 2D electron gas to generate an in-plane effective magnetic field $B_{BR}^{eff}$, which could be superimposed to the Zeeman field, facilitating the intravalley spin flip. In Fig. 1(a), we show theoretical calculation of the dependence of the magnitude of intrinsic spin splitting $|\lambda_{int}|$ on interlayer electric field $E_m$ for different conduction band interlayer hopping strength $t_\perp^c$ within a minimal band model of bilayer TMDs in the vicinity of the K points [9,17]. The insets show the configuration of the electric field perpendicular to the atomic plane (up) and electronic structure in the absence and presence of external electric field $E_m$ at the K point (down). With electric field applying, the inversion symmetry breaking lifts up spin degeneracy at Brillouin K points. We can see that $\lambda_{int}$ vanishes for $|E_m|=0$, suggesting no spin splitting in pristine bilayer MoS$_2$, in agreement with the symmetry argument. We found that in the limit of vanishing $t_\perp^c$, the spin splitting $|\lambda_{int}|$ saturates rapidly as soon as $E_m$ becomes non-zero and $|\lambda_{int}|$ increases slower with increasing $|E_m|$ as $t_\perp^c$ gets larger. In Fig. 1(b), we show theoretical calculation of the dependence of spin splitting $|\lambda_{int}|$ on interlayer electric field $E_m$ for different Bychkov-Rashba spin-orbit coupling strength $\lambda_{BR}$, within the same model. We found that for non-zero $\lambda_{BR}$, spin splitting $\lambda_{int}$ increases as $|E_m|$ gets larger. In this case, increasing $|E_m|$ can be understood as increasing in-plane effective magnetic field, thereby increasing total effective magnetic field. However, for this effect to be noticeable, the strength of $\lambda_{BR}$ has to be at least an order of magnitude larger than the value reported in Ref. [16]. Hence, the tunability of spin splitting requires either strong conduction (valence) band interlayer hopping strength for electrons (holes) or strong Bychkov-Rashba spin-orbit coupling. In both Fig. 1(a) and (b), we have set the carrier type to be electron, which is in accordance with our experiment. In our theoretical model, the range of $E_m$ has been estimated using self-consistent model of screening [17,18].

We now discuss the quantum coherent transport experimental results. The inset of Fig. 2(a) shows a schematic illustration of our double gate device with an ion-gel top gate and solid Si back gate. Details on the device preparation have been discussed in our previous work [19]. All of charge- and magneto-transport measurements were conducted in a Helium-4 cryostat. Figure 2(a) shows the back-gate charge transfer curves for 1L, 2L, and 3L MoS$_2$ devices at three different top gate voltages (0 V, 1 V, 1.7 V), respectively. The transfer curves are individually shifted horizontally to match their threshold voltage for each curve. $V_{bg}^{eff}$ is the effective back-gate voltage corresponding to the total density in the 2D systems[19]. The charge densities are calculated with a capacitor model according to $n = C_{ox}\left(V_{bg}^{eff} - V_0\right)/e$, where $C_{ox}$ is the gate capacitance, and $V_0 = -220.3$ V is determined by extrapolation of phase coherence length $L_\varphi$ to 0 nm [17]. This nearly linear dependence of $L_\varphi$ on $V_{bg}^{eff}$ works well when the channel conductivity is larger than the conductance quantum. The charge carrier concentration then can be estimated to vary between $0.9 \times 10^{13}$ and $4.7 \times 10^{13}$/cm$^2$ for the entire voltage range. The dashed lines in Figure 2(a) are fits to the conductivity according to $\sigma = \alpha\left(V_{bg}^{eff} - V_0\right)^\beta$ where $\alpha$ and $\beta$ are fitting parameters. $\beta$ is slightly larger than unity, indicating the superlinear dependence of $\sigma$ on $V_{bg}^{eff}$. For the data present in Fig 2(a), the electronic transport is diffusive, similar to our previous work on monolayer MoS$_2$ [20]. The mean free path $l$ and momentum relaxation time $\tau_p$ are therefore seen by $l = \upsilon_F \tau_p$ and $\tau_p = m^*\mu/e$ for mobility $\mu$, carrier density $n$, electron charge $e$ and effective mass $m^*$. The contact resistance is expected to play a minor role in this high doping regime due to the negligibly small depletion region [21].

For the double gate configuration, the average vertical electrical displacement field induced by combination of the two gates can be estimated with simple formula $D_m = \left(\kappa_{ox}V_b/d_b - \kappa_{ig}V_t/d_t\right)/2$ [22,23], with $d_t \approx 1$ nm and $d_b = 300$ nm the thickness of the top and bottom dielectric, $\kappa_{ox} \approx 3.9$ and $\kappa_{ig} \approx 7$ [24] the dielectric constant for SiO$_2$ and ion gel, and $V_t$ and $V_b$ the top and bottom gate voltages, respectively [23]. Here we assume the charge neutral point is at zero gate voltage. The direction of the external electric field is defined in inset to Fig. 1(a). Note that this formula doesn't take interlayer charge density imbalance into account. As seen in the figure, the electric displacement field falls in the range of about 4 V/nm, an overestimation of about an order of magnitude compared to the theoretical model employing self-consistent screening. This big discrepancy shows the importance of using the self-consistent screening theory to obtain the correct value of interlayer electric field in 2D parabolic system. This perpendicular electric field breaks the inversion symmetry of the bilayer system.

A perpendicular magnetic field breaks time reversal symmetry and leads to decoherence of the "closed-loop" current paths thereby suppressing WL, and generating a positive magnetoconductance $\Delta\sigma$. Figure 3 shows $\Delta\sigma$ as a function of both temperatures and gate voltages for 1~3L devices. The as-measured magnetoconductance curves were symmetrized to eliminate contributions from the sample geometry and Hall effect. Our analysis reveals that the magnetoresistance (MR) at high magnetic field is subjected to quadratic field dependent magnetoresistance containing both classical and quantum contributions. The pure quantum interference effect is extracted by subtracting the classical background [17]. As seen from the right panels of Figure 3, all devices exhibit a qualitatively similar trend characterized by positive $\Delta\sigma$ at low fields and a downward-turn at higher fields. This represents a crossover from WL to WAL with increasing fields, similar to our previous observation in monolayer MoS$_2$. The left side of each panel shows the theoretical fits, according to the Hikami-Larkin-Nagaoka (HLN) formula [25],

$$\Delta\sigma = g_l \frac{e^2}{\pi\hbar}\left[F\left(\frac{B}{B_\varphi}\right) - 3F\left(\frac{B}{B_\varphi + 2B_{SO}}\right)\right], \quad (1)$$

where $g_l$ represents the layer degeneracy, $F(Z)$ is defined by $F(z) = \ln(z) + \psi(1/2 + 1/z)$ with $\psi$ the digamma function, $B_\varphi = \hbar/4eL_\varphi^2$ and $B_{SO} = \hbar/4eL_{SO}^2$ are fitting parameters associated with the effective magnetic field for the phase coherence length and crossover length, respectively. We found that the formula with $g_l$ = 1, 2, 3 for monolayer, bilayer, and trilayer, respectively yields much better fittings of the experimental results.

We apply the same analysis from our previous work [20] in order to extract the spin lifetime from the WL-WAL crossover. Briefly, because of the large separation of length scales, the crossover can be attributed to symmetry breaking due to spin-flip scattering. The spin-conserved intervalley scattering is expected to break the valley-rotational symmetry due to high defect density in $MoS_2$, mainly in the form of chalcogen vacancies. By conducting a two-parameter fit to the temperature- and magnetic field-dependent magneto-conductance curves according to Eq. (1), we obtain the spin lifetime $\tau_{SO}$ from the relation $B_{SO} = \hbar/4eD\tau_{SO}$, where $D$ is the classical diffusion constant.

Figure 4 shows the ratio of spin relaxation time $\tau_{SO}$ divided by number of layers $g_l$ as a function of inverse momentum relaxation time $\tau_p^{-1}$, with inset showing a zoom in of the low spin relaxation time region. It can be seen that $\tau_{SO}$ scales linearly with $\tau_p^{-1}$ in the entirely range, indicating that the dominant spin relaxation mechanism in all the samples is D'yakonov-Perel (DP) type where spin precesses between scattering events from disorder impurities or non-magnetic defects. It is worth noting that bilayer $MoS_2$ exhibits the same DP behavior despite being inversion symmetric. From the slope of the fitted dashed line, the spin-orbit coupling strength $\lambda_{int}$ can be determined using the relationship $\tau_{SO} = \left(2\hbar^2/\lambda_{int}^2\right)\tau_p^{-1}$. The strength of spin-orbit interaction $\lambda_{int}$ of monolayer, bilayer and trilayer are found to be 7.2 meV, 6 meV, and 4.1 meV, respectively. These values represent the size of the conduction band splitting at the K point due to spin-orbit coupling. Similar trends for 1~3L devices and good agreement with Eq. (1) indicate that few-layer $MoS_2$ effectively behaves as $g_l$ decoupled monolayers.

Comparing with the theoretically predicted behavior (Figure 1), the absence of electrically tunable spin-orbit splitting energy in all 1~3L devices implies the following: First, the Bychkov-Rashba spin-orbit coupling strength $\lambda_{BR}$ is too weak to give rise to measurable changes in the spin-orbit interaction within the experimentally accessible electric field ranges. Second, a small perpendicular electric field is sufficient to break the inversion symmetry of bilayer $MoS_2$, lifting the spin degeneracy at the Brillouin Zone corner. Third, the interlayer hopping for electrons at the K points is negligible as expected from the symmetry argument [9]. Thus, once the inversion symmetry of bilayer is broken, it soon becomes two decoupled monolayers and tuning the spin-orbit interaction strength becomes difficult, which in turn explains the good fitting when including $g_l$ in Eq. (1).

## III. CONCLUSION

In this work, we have presented the coherent quantum electron transport properties of double-gated mono-, bi-, and trilayer MoS$_2$. We observed qualitatively similar crossover from WL to WAL in all the devices. The crossover can be explained by HLN formula with the spin-orbit scattering as symmetry breaking crossover parameter. Despite having different symmetries, all the samples exhibited spin relaxation due to DP mechanism. This implies that the breaking of inversion symmetry due to out-of-plane electric fields is not sufficient to tune the spin-relaxation rate. We conclude that this lack of electric-field induced changes in spin-orbit interaction strength implies weak Bychkov-Rashba spin-orbit coupling strength $\lambda_{BR}$ in TMD multilayers and a vanishing of interlayer coupling in the conduction band, which renders the bilayer system to behave like two decoupled monolayers. In contrast to conduction-band electrons, the interlayer hopping for valence band holes is finite (86 meV, Ref. [9]), and we therefore predict that a similar experiment perform instead on valence band holes would show a strong electric field tunability of the spin lifetimes.

**Notes added:**

After completion of this manuscript, we became aware of a similar work [26] that studied the phase-coherent transport properties of single-gated bilayer MoS$_2$ encapsulated in hexagonal boron nitride. The authors find $\tau_{SO}$ is almost constant with carrier density. In contrast to our results in dual-gated ion-gel devices, their data is not consistent with D'yakonov Perel spin-orbit relaxation. This discrepancy between our results and theirs is beyond the scope of this work.

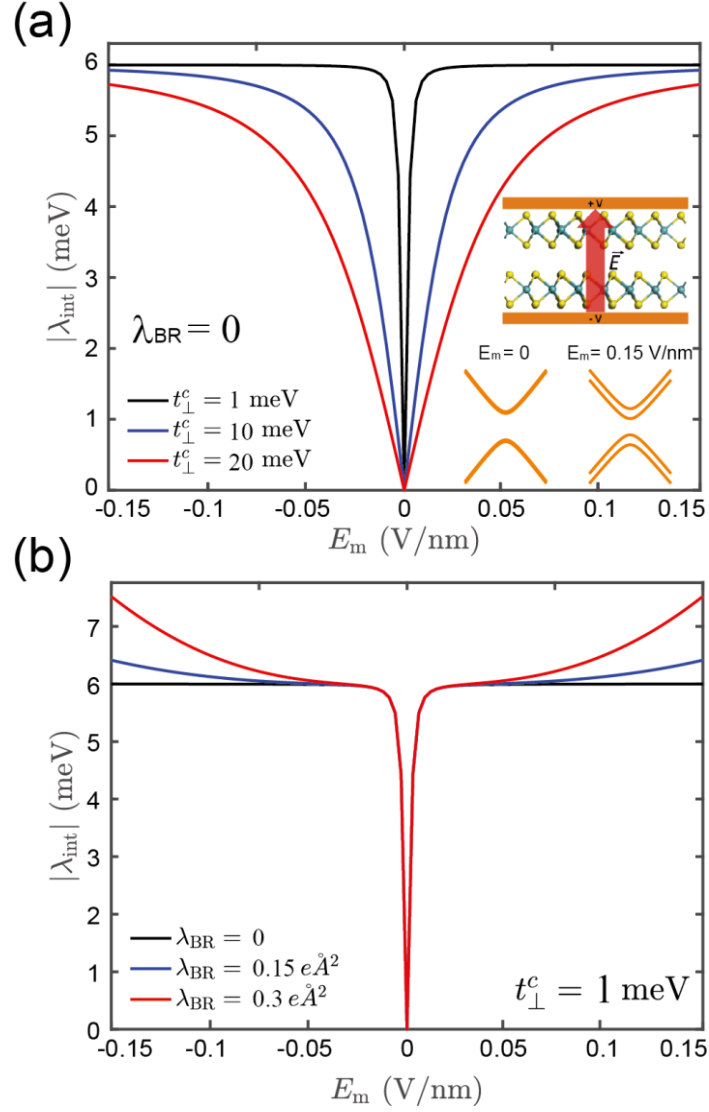

**FIG. 1.** Theoretical expectation of electric field tunability of spin lifetimes. (a) Theoretical calculation for the effect of conduction band interlayer coupling strength $t_\perp^c$ on the tunability of bilayer D'yakonov-Perel spin splitting $\lambda_{\text{int}}$ with an interlayer electric field $E_m$. At zero electric field, the spin splitting vanishes due to the presence of inversion symmetry and time-reversal symmetry. However, at large electric field, the inversion symmetry is broken, and the spin splitting approaches the single-layer value irrespective of coupling. Insets show a schematic illustration of the vertical

electric field through bilayer MoS$_2$ (up) and the K-point band structure at $E_m = 0$ and finite $E_m$ (down). (b) Theoretical calculation for the effect of Bychkov-Rashba spin-orbit coupling strength $\lambda_{BR}$ on the tunability of bilayer $\lambda_{int}$ with an interlayer coupling strength $t_\perp^c = 1$ meV. $\lambda_{int}$ increases with electric field when $\lambda_{BR}$ is non-zero, increasing more rapidly with larger $\lambda_{BR}$.

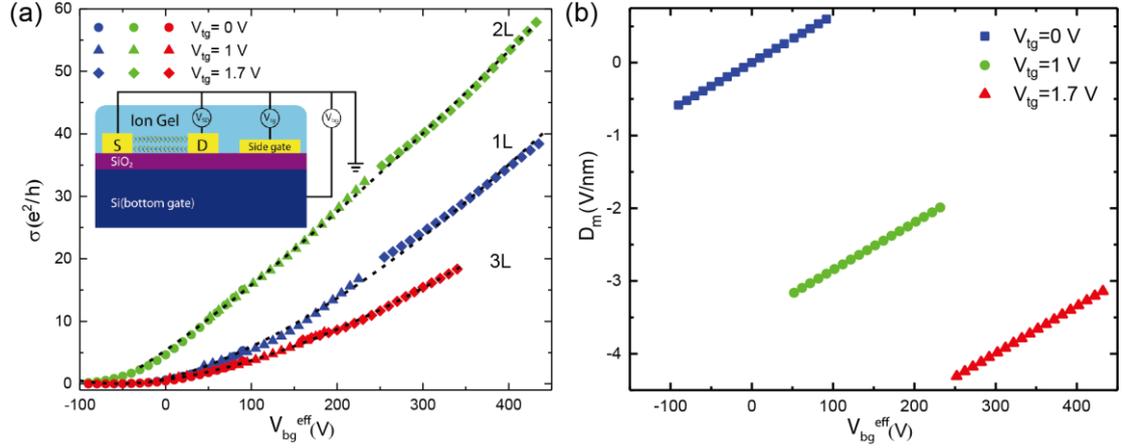

**FIG. 2.** (a) MoS$_2$ transfer characteristics for different layer thickness at various combinations of top gate and back-gate voltages. The transfer curves are shifted horizontally according to their conductivity with an indicated effective back-gate voltage bias. Inset shows the schematic illustration of the dual-gated device structure. The dashed lines are fitting curves. (b) Electrical displacement field strength at different combination of the dual gate voltages for simple model which ignores interlayer charge density imbalance.

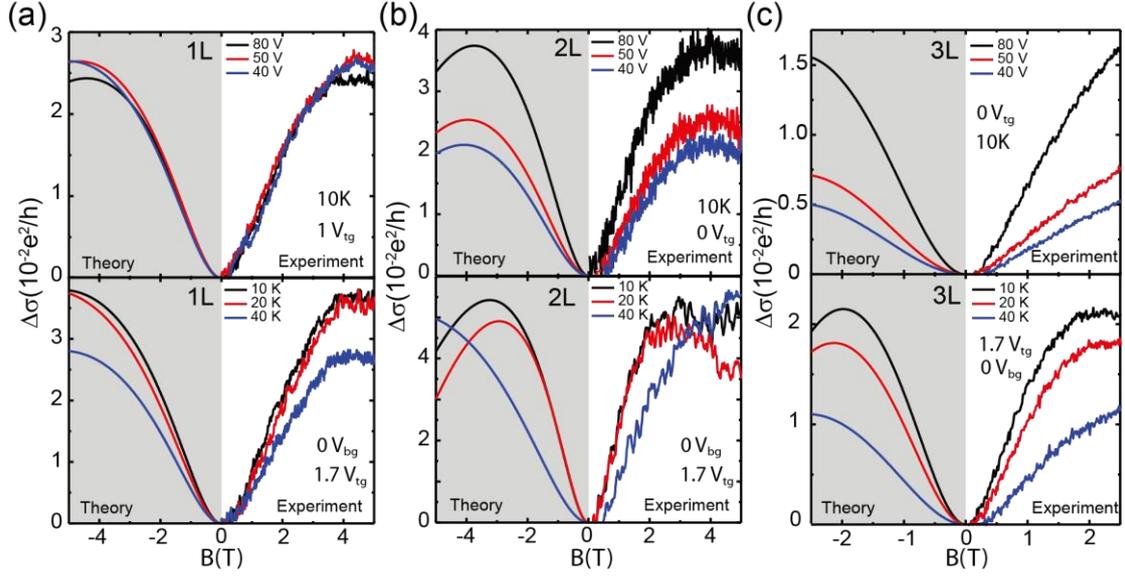

**FIG. 3.** (a-c) Top panel: Magneto-conductivity of monolayer, bilayer and trilayer MoS$_2$ at different back-gate voltages of 40, 50, and 80 V and fixed top gate voltage of $V_{tg}$ = 1 V(monolayer), $V_{tg}$ = 0 V(bilayer and trilayer) and fixed temperature of T = 10 K. Bottom panel: Magneto-conductivity of monolayer, bilayer and trilayer MoS$_2$ at different temperatures of 10, 20, and 40 K and fixed dual gate voltage of $V_{tg}$ = 1.7 V, $V_{bg}$ = 0 V. The left shaded curves are theoretically fits according to Eq. (1), while right curves in each figure are experimental results. To calculate $\Delta\sigma$, the measured magnetoresistance was symmetrized by $\rho(B) = [\rho(+B) + \rho(-B)]/2$, and that classical quadratic contribution has been subtracted.

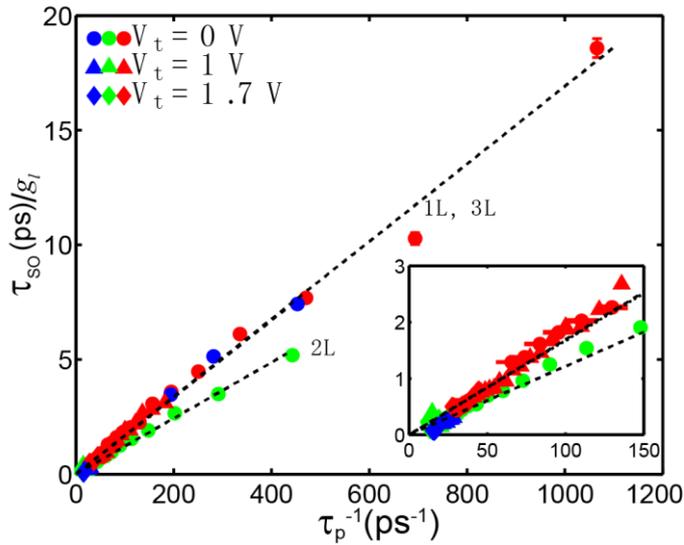

**FIG. 4.** D'yakonov-Perel spin relaxation in multilayer $MoS_2$ is electric field independent. The ratios of spin relaxation time $\tau_{SO}$ with number of layer $g_l$ as a function of inversed momentum relaxation time $\tau_p^{-1}$ correspond for monolayer, bilayer, and trilayer $MoS_2$, respectively. The dashed lines are fits using $2\hbar^2/\tau_p\tau_{SO} = \lambda_{int}^2$. The strength of spin-orbit interaction $\lambda_{int}$ are found to be 7.2 meV, 6 meV, and 4.1 meV, respectively. Inset shows a zoom in of the low spin relaxation time regime.

**ACKNOWLEDGEMENT**

We would like to thank Tang Shuai for assistance with the self-consistent screening calculation. G.E. acknowledges support from the Singapore Ministry of Education (MOE) under AcRF Tier 2 (MOE2015-T2-2-123, MOE2017-T2-1-134) and AcRF Tier 1 (R-144-000-387-114). S.A. acknowledges the National University of Singapore Young Investigator Award (R-607-000-094-133). L.C. acknowledges support from National Natural Science Foundation of China (Grant No. 11747169).

# Supporting Information for

# Phase coherent transport in bilayer and trilayer MoS$_2$

Leiqiang Chu,[1,2*] Indra Yudhistira,[1,3*] Hennrik Schmidt,[1,3] Tsz Chun Wu,[5] Shaffique Adam,[1,3,6,†] and Goki Eda[1,3,4,‡]

[1]*Department of Physics, National University of Singapore, 3 Science Drive 3, Singapore 117543*

[2]*Department of Physics, Shaoxing University, Shaoxing, China, 312000*

[3]*Centre for Advanced 2D Materials and Graphene Research Centre, National University of Singapore, 6 Science Drive 2, Singapore 117546*

[4]*Department of Chemistry, National University of Singapore, 2 Science Drive 3, Singapore 117551*

[5]*Department of Physics and Astronomy, Rice University, Houston, Texas 77005, USA*

[6]*Yale-NUS College, 6 College Ave West, Singapore 138614*

1. **Theoretical calculation for Figure 1a and 1b.**

A minimal model for bilayer MoS$_2$ can be constructed by adding interlayer hopping to the $\boldsymbol{k}\cdot\boldsymbol{p}$ model of monolayers[1],

$$H = \begin{bmatrix} -\frac{U}{2}+\frac{\Delta}{2}+\left(-s_z\frac{\lambda_c}{2}+E_1\frac{\lambda_{BR}}{2}\boldsymbol{s}\cdot\boldsymbol{k}\right) & atke^{i\theta_k} & t_\perp^c & 0 \\ atke^{-i\theta_k} & -\frac{U}{2}-\frac{\Delta}{2}+\left(-s_z\frac{\lambda_v}{2}+E_1\frac{\lambda_{BR}}{2}\boldsymbol{s}\cdot\boldsymbol{k}\right) & 0 & t_\perp^v \\ t_\perp^c & 0 & \frac{U}{2}+\frac{\Delta}{2}+\left(s_z\frac{\lambda_c}{2}+E_2\frac{\lambda_{BR}}{2}\boldsymbol{s}\cdot\boldsymbol{k}\right) & atke^{-i\theta_k} \\ 0 & t_\perp^v & atke^{i\theta_k} & \frac{U}{2}-\frac{\Delta}{2}+\left(s_z\frac{\lambda_v}{2}+E_2\frac{\lambda_{BR}}{2}\boldsymbol{s}\cdot\boldsymbol{k}\right) \end{bmatrix}$$

(S1)

where $\Delta$ is bandgap of monolayer MoS$_2$, $\lambda_c(\lambda_v)$ is the intrinsic spin-orbit coupling (SOC) strength for conduction (valence) band, $\lambda_{BR}$ is Bychkov-Rashba SOC strength, $U$ is interlayer potential, given by

$$U = eE_m d_m, \text{(S2)}$$

with $E_m$ and $d_m$ are interlayer electric field and interlayer distance, respectively, while $E_1$ ($E_2$) is the electric field in the bottom (top) layer, given by

$$E_{1(2)} = \frac{E_m + E_{b(t)}}{2}, (S3)$$

where $E_b$ ($E_t$) are the electric field inside the bottom (top) gate.

We projected the hamiltonian into 2x2 effective Hamiltonian of the conduction band, since in our experiment the sample is electron doped.

$$H_{eff} = \frac{\Delta}{2} + \begin{bmatrix} \frac{\hbar^2 k^2}{2m_1} - \frac{U}{2} - s_z \frac{\lambda_{eff}^{(1)}(k)}{2} & \bar{t}_\perp(k) \\ \bar{t}_\perp(k) & \frac{\hbar^2 k^2}{2m_2} + \frac{U}{2} + s_z \frac{\lambda_{eff}^{(2)}(k)}{2} \end{bmatrix}, (S4)$$

where

$$m_{1(2)} = \frac{\hbar^2}{2a^2 t^2} \frac{\left(\frac{\Delta}{2}\right)^2 - \left(\frac{U}{2}\right)^2 - (t_\perp^v)^2}{\frac{\Delta}{2} \mp \frac{U}{2}}$$

$$\bar{t}_\perp(k) = t_\perp^c + \frac{(atk)^2}{\left(\frac{\Delta}{2}\right)^2 - \left(\frac{U}{2}\right)^2 - (t_\perp^v)^2} t_\perp^v$$

$$\lambda_{eff}^{1(2)} = \sqrt{\lambda_c^2 + (E_{1(2)} \lambda_{BR} k)^2}$$

The energy eigenvalue of the conduction band is then given by

$$E_{\pm, s_z}(k, s_z) = \frac{\Delta}{2} + (atk)^2 \frac{\frac{\Delta}{2}}{\left(\frac{\Delta}{2}\right)^2 - \left(\frac{U}{2}\right)^2 - (t_\perp^v)^2} + s_z \frac{\lambda_{eff}^{(2)}(k) - \lambda_{eff}^{(1)}(k)}{4}$$

$$\pm \sqrt{\left[\left(1 + \frac{(atk)^2}{\left(\frac{\Delta}{2}\right)^2 - \left(\frac{U}{2}\right)^2 - (t_\perp^v)^2}\right) \frac{U}{2} + s_z \frac{\lambda_{eff}^{(2)}(k) - \lambda_{eff}^{(1)}(k)}{4}\right]^2 + \bar{t}_\perp^2(k)}, (S5)$$

where $\pm$ refers to upper and lower band and $s_z$ for spin up/down.

The spin splitting for the upper and lower band $\lambda_{int}^\pm$ is then given by

$$\lambda_{int}^{\pm} = E_{\pm}(s_z = 1) - E_{\pm}(s_z = -1)$$

$$= (at)^2\left(k_{\pm,\uparrow}^2 - k_{\pm,\downarrow}^2\right)\frac{\frac{\Delta}{2}}{\left(\frac{\Delta}{2}\right)^2 - \left(\frac{U}{2}\right)^2 - (t_{\perp}^v)^2} + \frac{\lambda_{eff}^{(2)}(k_{\pm,\uparrow}) - \lambda_{eff}^{(1)}(k_{\pm,\uparrow})}{4}$$

$$+ \frac{\lambda_{eff}^{(2)}(k_{\pm,\downarrow}) - \lambda_{eff}^{(1)}(k_{\pm,\downarrow})}{4}$$

$$\pm \left\{ \sqrt{\left[\left(1 + \frac{(atk_{\pm,\uparrow})^2}{\left(\frac{\Delta}{2}\right)^2 - \left(\frac{U}{2}\right)^2 - (t_{\perp}^v)^2}\right)\frac{U}{2} + s_z\frac{\lambda_{eff}^{(2)}(k_{\pm,\uparrow}) - \lambda_{eff}^{(1)}(k_{\pm,\uparrow})}{4}\right]^2 + \bar{t}_{\perp}^2(k_{\pm,\uparrow})} \right.$$

$$\left. - \sqrt{\left[\left(1 + \frac{(atk_{\pm,\downarrow})^2}{\left(\frac{\Delta}{2}\right)^2 - \left(\frac{U}{2}\right)^2 - (t_{\perp}^v)^2}\right)\frac{U}{2} + s_z\frac{\lambda_{eff}^{(2)}(k_{\pm,\downarrow}) - \lambda_{eff}^{(1)}(k_{\pm,\downarrow})}{4}\right]^2 + \bar{t}_{\perp}^2(k_{\pm,\downarrow})} \right\}, (S6)$$

Approximating $k_{\pm,\uparrow} \approx k_{\pm,\downarrow} \equiv k_{\pm}$, we get

$$\lambda_{int}^{\pm} = \frac{\lambda_{eff}^{(2)}(k_{\pm}) - \lambda_{eff}^{(1)}(k_{\pm})}{2}$$

$$\pm \left\{ \sqrt{\left[\left(1 + \frac{(atk_{\pm})^2}{\left(\frac{\Delta}{2}\right)^2 - \left(\frac{U}{2}\right)^2 - (t_{\perp}^v)^2}\right)\frac{U}{2} + \frac{\lambda_{eff}^{(2)}(k_{\pm}, E_m) - \lambda_{eff}^{(1)}(k_{\pm}, E_m)}{4}\right]^2 + \bar{t}_{\perp}^2(k_{\pm}, E_m)} \right.$$

$$\left. - \sqrt{\left[\left(1 + \frac{(atk_{\pm})^2}{\left(\frac{\Delta}{2}\right)^2 - \left(\frac{U}{2}\right)^2 - (t_{\perp}^v)^2}\right)\frac{U}{2} - \frac{\lambda_{eff}^{(2)}(k_{\pm}, E_m) - \lambda_{eff}^{(1)}(k_{\pm}, E_m)}{4}\right]^2 + \bar{t}_{\perp}^2(k_{\pm}, E_m)} \right\}, (S7)$$

The interlayer electric field $E_m$ is calculated following the screening model of McCann *et al*[2]. We assume that bilayer MoS$_2$ consists of two parallel conducting plates separated by interlayer distance $d_m$ and dielectric constant $\kappa_m$ (see Fig. S1). Each layers supports electron densities of $n_1$ ($n_2$) for bottom (top) layer, with corresponding back (top) gate held at potential $V_b$ ($V_t$) with dielectric constant $\varepsilon_b$ ($\varepsilon_t$), having thickness of $d_b$ ($d_t$). Meanwhile, the dielectric constant of bilayer interlayer space is denoted by $\kappa_m$.

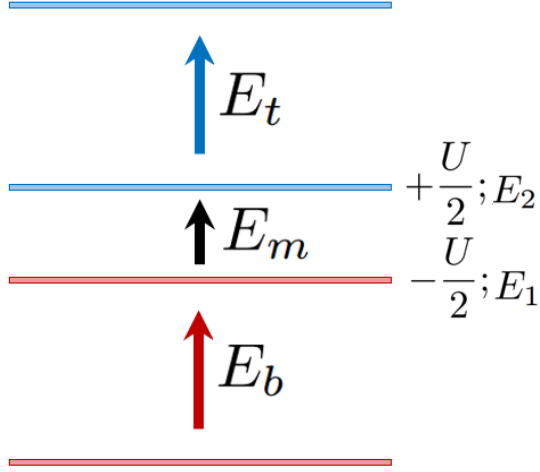

**FIG. S1**. Schematic of bilayer MoS$_2$ in the presence of back and top gates.

Application of Gauss' law to the Gaussian surface having cross-sectional area $A$ formed by enclosing both layers, as well as one layer only, yields

$$\varepsilon_0(-\kappa_b E_b + \kappa_t E_t)A = -e(n_1 + n_2)A, \quad (S8a)$$

$$\varepsilon_0(-\kappa_m E_m + \kappa_t E_t)A = -en_2 A, \quad (S8b)$$

respectively, where $E_b$ ($E_t$) are the electric field inside the bottom (top) gate.

Estimating the electric field of the bottom and top gate as

$$E_{b(t)} \approx \pm \frac{V_{b(t)}}{d_{b(t)}}, \quad (S9)$$

and substitute them to Eq. (S8), we obtain total charge density $n$ and interlayer electric field $E_m$,

$$n = n_1 + n_2 = \frac{\varepsilon_0}{e}\left(\frac{\kappa_b V_b}{d_b} + \frac{\kappa_t V_t}{d_t}\right), \quad (S10a)$$

$$E_m = \frac{1}{2d_m}\left(\frac{\alpha_b}{\alpha_m}V_b - \frac{\alpha_t}{\alpha_m}V_t + \frac{\Delta n}{\alpha_m}\right), \quad (S10b)$$

where $\Delta n \equiv n_2 - n_1$ is the charge density imbalance between layers, and $\alpha_j$ is capacitance per area per electron, given by

$$\alpha_j \equiv \frac{C_j}{eA} = \frac{\varepsilon_0 \kappa_j}{ed_j}.$$

Here, subscript $j$ are either $t, b,$ or $m$ for bottom, top, and middle, respectively.

## 1.1 Non-self-consistent calculation

The simplest estimation of interlayer electric field $E_m$ can be obtained by neglecting interlayer density imbalance ($\Delta n$) contribution. This yields

$$E_m \approx \frac{1}{2d_m}\left(\frac{\alpha_b}{\alpha_m}V_b - \frac{\alpha_t}{\alpha_m}V_t\right)$$

$$= \frac{1}{2\kappa_m}\left(\kappa_b \frac{V_b}{d_b} - \kappa_t \frac{V_t}{d_t}\right). \quad (S11)$$

This widely used expression overestimates the interlayer electric field by about an order of magnitude, i.e. ~30 times compared to the result of self-consistent calculation [3]. This observation has been considered when we estimate the range of interlayer electric field in Fig. 1a and 1b of the main text.

## 1.2 Self-consistent calculation

In order to obtain a good quantitative value of interlayer electric field $E_m$, one need to take the charge density imbalance ($\Delta n$) contribution, and solve the equation self-consistently. To obtain the interlayer charge density imbalance $\Delta n$, the charge density in each layer $n_{1(2)}$ is calculated by integration over Fermi surface taking into account the probability of occupancy in each layer

$$n_1^{c\pm} = 4\int_0^{k_F^\pm} \frac{d^2k}{(2\pi)^2}\left(\left|\psi_1^{c,\pm}(k,E_m)\right|^2 + \left|\psi_2^{c,\pm}(k,E_m)\right|^2\right), \quad (S12)$$

$$n_2^{c\pm} = 4\int_0^{k_F^\pm} \frac{d^2k}{(2\pi)^2}\left(\left|\psi_3^{c,\pm}(k,E_m)\right|^2 + \left|\psi_4^{c,\pm}(k,E_m)\right|^2\right)$$

$$n_1^{v\pm} = 4\int_0^{k_{max}} \frac{d^2k}{(2\pi)^2}\left(\left|\psi_1^{v,\pm}(k,E_m)\right|^2 + \left|\psi_2^{v,\pm}(k,E_m)\right|^2\right)$$

$$- 4\int_0^{k_{max}} \frac{d^2k}{(2\pi)^2}\left(\left|\psi_1^{v,\pm}(k,E_m=0)\right|^2 + \left|\psi_2^{v,\pm}(k,E_m=0)\right|^2\right)$$

$$n_2^{v\pm} = 4\int_0^{k_{max}} \frac{d^2k}{(2\pi)^2}\left(\left|\psi_3^{v,\pm}(k,E_m)\right|^2 + \left|\psi_4^{v,\pm}(k,E_m)\right|^2\right)$$

$$- 4\int_0^{k_{max}} \frac{d^2k}{(2\pi)^2}\left(\left|\psi_3^{v,\pm}(k,E_m=0)\right|^2 + \left|\psi_4^{v,\pm}(k,E_m=0)\right|^2\right),$$

where the superscript $c(v)$ refers to conduction (valence) band, the $\pm$ sign refers to the higher (lower) band, $\psi_j^{c(v),\pm}$ with $j$=1,2,3,4 are components of eigenvector $\Psi^{c(v),\pm} = \left(\psi_1^{c(v),\pm}\ \psi_2^{c(v),\pm}\ \psi_3^{c(v),\pm}\ \psi_4^{c(v),\pm}\right)^T$, and $k_{max} \sim O(2\pi/a)$ is ultraviolet cutoff. Here $a$ is size of unit cell.

For this interlayer electric field ($E_m$) calculation, we have used the original 4x4 effective Hamiltonian of Eq. (S1) with $\lambda_c \approx \lambda_{BR} \approx 0$ approximation.

Combining contribution from both conduction and valence band, we obtain individual layer density

$$n_{1(2)} = \begin{cases} n_{1(2)}^{c-} + n_{1(2)}^{v-} + n_{1(2)}^{v+} & ; if\ E_F < \Delta/2 + \sqrt{t_c^2 + (U/2)^2} \\ n_{1(2)}^{c-} + n_{1(2)}^{c+} + n_{1(2)}^{v-} + n_{1(2)}^{v+} & ; otherwise \end{cases}, (S13)$$

Here, we have assumed that $E_F$ lies in conduction band, in accordance with the condition in our experiment.

Interlayer electric field $E_m$ is obtained by solving self-consistently Eq. (S10b), with Δn calculated using Eq. (S12) and Eq. (S13).

Finally, one needs these equations to close the self-consistency equation.

$$\begin{cases} k_F^- = \sqrt{\pi n} & ; if\ E_F < \Delta/2 + \sqrt{t_c^2 + (U/2)^2} \\ (k_F^-)^2 + (k_F^+)^2 = \pi n & \\ E_-(k_F^-, E_m) = E_+(k_F^+, E_m) & ; otherwise \end{cases}, (S14)$$

2. **Determination of $V_0$.**

To determine $V_0$, we first extract the phase coherence length $L_\varphi$ by fitting the quantum transport data with HLN theory. We then plot it as a function of the effective back-gate voltage $V_{bg}^{eff}$, as shown in Figure S2. Finally, $V_0$ is obtained from linear extrapolation to $L_\varphi = 0$.

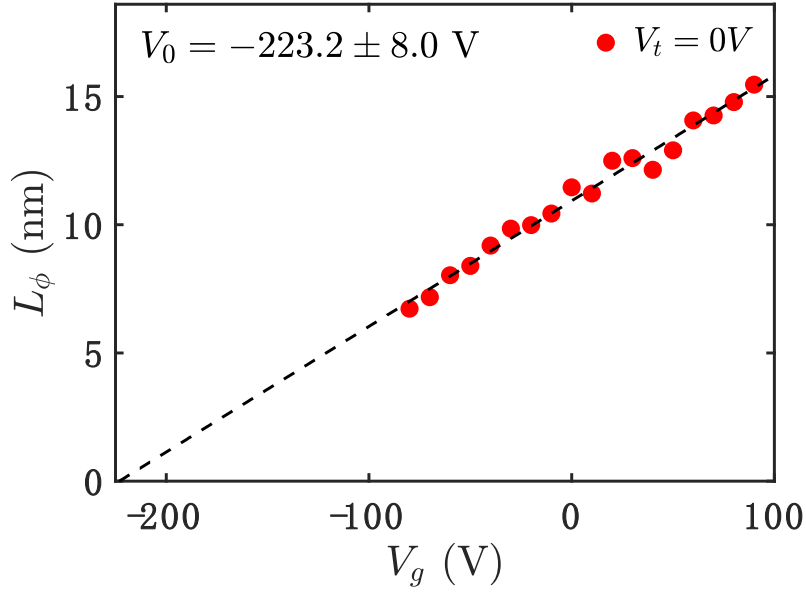

**FIG. S2**. Phase coherence length of bilayer MoS$_2$ as a function of effective back gate voltage.

### 3. Procedure for extract the pure quantum interference effect.

The magnetoresistance(MR) is defined as:

$$MR(B) = \frac{R(B) - R(0)}{R(0)} \times 100\%, (S15)$$

where *R(B)* is the sheet resistance at a given magnetic field B.

Ordinary magnetoresistance (OMR) is a classical effect due to cyclic motion of the electrons. It typically follows kohler's rule:

$$MR \sim (\omega_c \tau)^2 = (\mu B)^2, (S16)$$

where $\omega_c$ is the cyclotron frequency, $\tau$ is the momentum scattering time, and *μ is* the carrier mobility. As shown in Figure S3, a quadratic fitting collapse well with the MR at high magnetic field, indicating our MR at high field contains contributions from both classical and quantum interference effects. We subtract the classical background to reveal the pure quantum correction as shown in the inset red curve. To further verify our assumption of the OMR at high field, we plot MR as a function with B$^2$. As shown in Fig. S1b, the data collapse well on a single linear curve for B > 4 T for various gate-voltage combinations. The curves are normalized according to the quadratic fitting to show their consistent linear behaviour at high magnetic field.

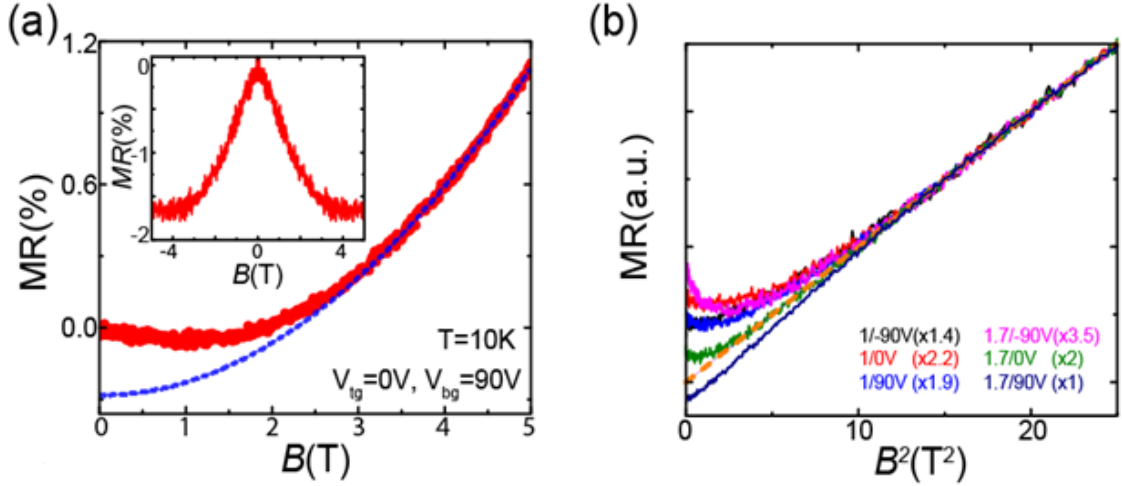

**FIG. S3.** Classical and quantum corrections. (a) Magnetoresistance at high field contains a quadratic classical contribution, as indicated by the fitting dashed blue line. The inset shows the calculated contribution from pure quantum interference effect after subtraction of the classical background. In the main text, the magneto-transport data shown are all pure quantum contributions. (b) MR scales as a function of $B^2$ for several combinations of dual gate voltages. The curves are all normalized by multiplying factors (shown in the brackets). As indicated by the orange dashed line, it is apparently linear for all curves at high magnetic fields.